\documentclass[fleqn,usenatbib]{mnras}

\usepackage[T1]{fontenc}

\DeclareRobustCommand{\VAN}[3]{#2}
\let\VANthebibliography\thebibliography
\def\thebibliography{\DeclareRobustCommand{\VAN}[3]{##3}\VANthebibliography}

\usepackage{graphicx}	
\usepackage{amsmath}	
\usepackage{amssymb}

\usepackage{newtxtext,newtxmath}
\usepackage{lipsum}
\usepackage{xspace}
\usepackage{pdflscape}
\usepackage[dvipsnames]{xcolor}

\newcommand{\man}{MaNGA\xspace}
\newcommand{\sam}{SAMI\xspace}

\newcommand{\dd}{{\rm d}}

\newcommand{\pipe}{\textsc{pyPipe3D}\xspace}
\newcommand{\ppxf}{\textsc{PPXF}\xspace}

\newcommand{\angstrom}{\text{\normalfont\AA}\xspace}
\newcommand{\Msun}{\ensuremath{\rm M_\odot}\xspace}

\newcommand{\ssfr}{\ensuremath{sSFR}\xspace}
\newcommand{\ha}{\ensuremath{\rm H\alpha}\xspace}
\newcommand{\ew}{\ensuremath{\rm EW(\ha)}\xspace}
\newcommand{\reff}[1]{\ensuremath{#1\,R_{\rm e}}\xspace}
\newcommand{\balbreak}{\ensuremath{D_n(4000)}\xspace}

\newcommand{\mass}{\ensuremath{M_\star}}
\newcommand{\logmass}{\ensuremath{\log_{10}(\mass/{\rm M_\odot})}\xspace}
\newcommand{\re}{\ensuremath{R_{e}}\xspace}
\newcommand{\rfifth}{\ensuremath{R_{50}}\xspace}
\newcommand{\rpetro}[1]{\ensuremath{R_{#1}}\xspace}
\newcommand{\logre}{\ensuremath{\log_{10}(\rfifth/ \rm kpc)}\xspace}
\newcommand{\concen}{\ensuremath{\frac{\rpetro{90}}{\rpetro{50}}}\xspace}
\newcommand{\logz}{\ensuremath{\log_{10}(Z_\star/{\rm Z_\odot})}\xspace}
\newcommand{\vsigma}{\ensuremath{v/\sigma}\xspace}
\newcommand{\envir}{\ensuremath{\Sigma_5}\xspace}
\newcommand{\logenvir}{\ensuremath{\log_{10}(\envir/\rm Mpc^{-2})}\xspace}
\newcommand{\halfrad}{\ensuremath{r_{50}}\xspace}
\newcommand{\totrad}{\ensuremath{r_{90}}\xspace}

\newcommand{\change}[1]{#1}

\defcitealias{Casado+15}{C15}
\defcitealias{Corcho-Caballero+22}{CC23}
\defcitealias{Corcho-Caballero+21a}{CC21a}
\defcitealias{Corcho-Caballero+21b}{CC21b}
\defcitealias{Corcho-Caballero+20}{CC20}

\title[Ageing and Quenching through the ageing diagram II]{Ageing and Quenching through the ageing diagram II: physical characterization of galaxies.}

\author[P. Corcho-Caballero et al.]{
	Pablo Corcho-Caballero,$^{1,2,3}$\thanks{E-mail: pablo.corcho@uam.es}
	Yago Ascasibar,$^{1}$
	Luca Cortese,$^{3, 4}$
	Sebasti\'an F. S\'anchez,$^{5}$	
	\newauthor
	\'Angel R. L\'opez-S\'anchez$^{2,4,6}$,
	Amelia Fraser-McKelvie,$^{3, 4}$
	and Tayyaba Zafar$^{2,4,6}$
	\\
	$^{1}$Departamento de Física Teórica, Universidad Autónoma de Madrid (UAM), Campus de Cantoblanco, Madrid 28049, Spain\\
	$^{2}$Australian Astronomical Optics, Macquarie University, 105 Delhi Rd, North Ryde, NSW 2113, Australia\\
	$^{3}$ARC Centre of Excellence for All Sky Astrophysics in 3 Dimensions (ASTRO-3D)\\
	$^{4}$International Centre for Radio Astronomy Research, The University of Western Australia, 35 Stirling Hwy., 6009 Crawley, WA, Australia\\
	$^{5}$Instituto de astronom\'ia, Universidad Nacional Aut\'onoma de M\'exico, A.P. 70-264, 04510 M\'exico D. F., M\'exico\\
	$^{6}$Macquarie University Research Centre for Astronomy, Astrophysics \& Astrophotonics, Sydney, NSW 2109, Australia.
}

\date{Accepted XXX. Received YYY; in original form ZZZ}

\pubyear{2021}

\begin{document}
	\label{firstpage}
	\pagerange{\pageref{firstpage}--\pageref{lastpage}}
	\maketitle
	
	\begin{abstract}
		The connection between quenching mechanisms, which rapidly turn star-forming systems into quiescent, and the properties of the galaxy population remains difficult to discern. In this work we investigate the physical properties of MaNGA and SAMI galaxies at different stages of their star formation history. Specifically, we compare galaxies with signatures of recent quenching (Quenched) \change{-- \ha in absorption and low \balbreak --} with the rest of the low star-forming and active population (Retired and Ageing, respectively). The analysis is performed in terms of characteristics such as the total stellar mass, half-light radius, velocity-to-dispersion ratio, metallicity, and environment. We find that the Ageing population comprises a heterogeneous mixture of galaxies, preferentially late-type systems, with diverse physical properties. Retired galaxies, formerly Ageing or Quenched systems, are dominated by early-type high-mass galaxies found both at low and dense environments. Most importantly, we find that recently quenched galaxies are consistent with a population of compact low-mass satellite systems, with higher metallicities than their Ageing analogues. We argue that this is compatible with being quenched after undergoing a star-burst phase induced by environmental processes (e.g. ram pressure). However, we also detect a non-negligible fraction of field central galaxies likely quenched by internal processes. This study highlights that, in order to constrain the mechanisms driving galaxy evolution, it is crucial to distinguish between old (Retired) and recently quenched galaxies, thus requiring at least two estimates of the specific star formation rate over different timescales.
	\end{abstract}
	
	\begin{keywords}
		galaxies: star formation -- galaxies: evolution -- galaxies:stellar content -- galaxies: general
	\end{keywords}
	
	
	
	\section{Introduction}
	\label{sec:intro}
	
	The connection between the star formation history (SFH) of galaxies and their physical properties (e.g. total stellar mass, chemical composition, morphology, kinematics, environment) has been extensively studied during the last few decades \citep{Kauffmann+03, Tremonti+04, Gallazzi+05, Peng+10}.
	However, due to the complex mixture of physical processes acting upon galaxies, drawing a complete picture of galaxy evolution is still very challenging.
	
	For a number of physical properties the statistical distribution of galaxies appears to be bimodal, e.g. light concentration \citep{Strateva+01} or disk vs spheroidal morphology \citep{Wuyts+11}, UV to optical colours and absorption line features \citep{Baldry+04, Gallazzi+05}, or internal kinematics \citep{Emsellem+11}.
	This dichotomy has also been translated into the plane formed between the total stellar mass, \mass, and the specific star formation rate (sSFR, where $\ssfr \equiv \frac{SFR}{\mass}$) \citep[e.g.][]{Salim+07, Renzini&Peng15, Belfiore+17}.
	Galaxies are divided into those along the so-called Main Sequence of star formation (MS) \citep{Noeske+07}, i.e. a tight relation between the total stellar mass and the star formation rate (SFR), and a detached population of ``passive'' galaxies scattering below the MS with lower levels of star formation.
	Nevertheless, several recent studies showed that the bimodal interpretation (in terms of several properties) might be too simplistic \citep{Casado+15, Abramson+16, Eales+18b, Corcho-Caballero+20, Fraser-McKelvie+22}.
    While the fact that galaxies form a bimodal population in some parameter space could imply a common evolutionary path, it is also possible that this dichotomy is driven by the lack of sensitivity of the observables, unable to capture the diversity of mass assembly histories.
	
	The bimodal interpretation poses the existence of some ``quenching'' processes able to terminate star formation in a galaxy in multiple ways, converting star-forming (blue) galaxies into passive (red) ones.
	During the last years, a large number of physical processes have been proposed as potential agents that shut down star formation.
	On the one hand, galaxies might quench due to environmentally-driven processes such as ram pressure stripping \citep[able to remove the gas reservoir, e.g.][]{Gunn&Gott72, Boselli&Gavazzi06, Brown+17}, strangulation/starvation \citep[leading to the suppression of gas infall, e.g.][]{Wetzel+13, Peng+15} or galaxy interactions \citep[][]{Moore+96, Bialas+15}, see \citet{Cortese+21}, for a review.
	On the other hand, internally-triggered quenching mechanisms such as negative feedback from Active Galactic Nuclei (AGN), supernovae driven winds \citep[][]{Croton+06, Sawala+10, Cheung+16, Fitts+17} or the stabilization of the gas against fragmentation \change{\citep[e.g.][]{Bigiel+08, Martig+09, Gensior+20}}, might also cause the demise of star formation in galaxies.
	
	It is important to remark that, although most of these processes are thought to bring star formation to a halt, they can also produce the opposite effect; to enhance star formation efficiency and lead to the depletion of the gas reservoir in very short timescales \citep[][]{Zinn+13, Cresci+15, Thorp+22}.
	Moreover, the term ``quenching'' is loosely defined, and different meanings can be found in the literature.
	In this work, we will refer to as \emph{quenching} as a process able to terminate\footnote{or significantly suppress, provided a galaxy never completely shuts off its star formation  \citep[see][]{Corcho-Caballero+21a}.} the star formation processes of a galaxy in a short timescale compared to the age of the Universe (i.e., $\lesssim$ 1 Gyr).
	In contrast, we use the term \emph{ageing} to denote the continuous evolution of a galaxy driven by the consumption of the gas reservoir -- encompassing different evolutionary stages from starburst to quiescent phases -- through uninterrupted star formation.
	The ageing scenario implies that all galaxies eventually become dominated by old stellar populations, turning red, without the need of a particular event that truncates star formation \citep[e.g.][]{Abramson+16}.
	\change{Unfortunately, discriminating between ageing galaxies and systems whose quenching timescales are larger than $\sim 1$ Gyr, often referred to as ``slow quenching'' \citep[e.g.][]{Belli+19, Tacchella+22}, is in practice extremely challenging, if not impossible, and therefore we will focus on galaxies that show evidence of recent ``fast quenching''.}

	There are numerous examples in the literature of attempts to identify quenching in the Universe.
	Some previous studies employed a combination of UV to IR photometric measurements to provide a division in terms of star-forming and quenched galaxies \citep[e.g.][]{Williams+09, Schawinski+14}.
	We would like to argue that this method does not offer a clean separation between ``star forming'' and ``quenched galaxies'' unless one assumes that all red systems were in fact quenched \citep[see e.g.][for an alternative interpretation]{Abramson+16}.
	\change{Closer to our approach, there have recently been several attempts to characterise the time derivative of the SFH in nearby galaxies \citep{Martin+17, Wang+20, Jimenez-Lopez+22, Weibel+22}.
    These works built large sets of mock SFHs to derive synthetic observables, such as broad-band colours, and/or spectral features like \ew, $\rm EW(H\delta)$, and \balbreak.
    Then, multi-dimensional polynomials and different regression techniques were employed to infer the average star formation rate over different timescales.
    }
    	
	\change{
    Building upon several observational samples of nearby galaxies ($z\lesssim 0.1$), we developed in previous works a complementary approach: instead of deriving quantitative estimates of recent changes in the SFR, we proposed an empirical} diagnostic diagram to discriminate between fast and slow evolution \citep[][hereafter \citetalias{Casado+15}, \citetalias{Corcho-Caballero+21b}, and \citetalias{Corcho-Caballero+22}, respectively]{Casado+15, Corcho-Caballero+21b, Corcho-Caballero+22}.
	The \emph{Ageing Diagram} (AD) combines two proxies for star formation, sensitive to different timescales, to probe the derivative of the recent SFH of galaxies during the last $\sim1-3$ Gyr.
	We use the raw \ew~-- including both absorption and emission components -- to trace the recent star formation during the last $\sim10^{6}-10^{7}$ yr, while we employ optical colours \citepalias[$u-r$ and $g-r$ in][respectively]{Casado+15, Corcho-Caballero+21b} or \balbreak \citepalias{Corcho-Caballero+22} to trace the sSFR over the last $\sim10^9$ yr.
	Systems whose SFR varies smoothly over time will arrange along a sequence given by a tight correlation between both proxies.
	In contrast, galaxies that experienced recent quenching episodes will feature significantly smaller values of \ew, due to the dearth of O and B stars, while still displaying a stellar continuum dominated by intermediate stellar populations (roughly corresponding to A-type stars).
	
	This method is also similar to the classical approach adopted for selecting post-starburst galaxies \citep[PSBGs, e.g.][]{Dressler&Gunn83, Poggianti+99, Balogh+05, Owers+19, Chen+19, Zheng+20}, systems with prominent absorption Balmer lines ($\rm EW(H\delta)< 5~\angstrom$, strongly correlated with \balbreak) and mild to null nebular emission.
	Therefore, under our prescription, PSBGs would roughly correspond to the blue end of the quenched sequence.

    \change{
    As noted in \citetalias{Casado+15}, our qualitative classification is complementary to the numerical estimation of the time derivative of the (NUV-i) colour \citep{Martin+17} or SFR \citep[e.g.][]{Jimenez-Lopez+22}, or the ratio between the SFR averaged between different timescales \citep[e.g.][]{Corcho-Caballero+22, Weibel+22}).
	}
	In \citetalias{Casado+15}, we performed a statistical analysis of SDSS fibre spectra and showed that the vast majority of galaxies lie across the \emph{ageing sequence} -- consistent with smoothly declining SFHs spanning a wide range of sSFR values.
	The spatially resolved distribution of CALIFA galaxies across the AD was studied in \citetalias{Corcho-Caballero+21b}, where we found that only a handful of low-mass early-type galaxies were fully quenched but the reduced number of systems in this range precluded from any robust conclusion.
	Finally, in \citetalias{Corcho-Caballero+22} we studied the distribution of CALIFA, MaNGA, and IllustrisTNG galaxies across the AD using their integrated spectra within one effective radius, and its connection with their star formation histories and evolutionary timescales.
	We proposed two demarcation lines used to classify galaxies into four AD domains: Ageing galaxies (AGs); systems undergoing secular evolution, Undetermined galaxies (UGs); an intermediate class populated with galaxies with unclear classification, Quenched galaxies (QGs); systems that experienced some quenching event during the last $\sim 1$ Gyr, and Retired galaxies (RGs); located at the red end where ageing and quenched sequences converge, whose recent past becomes extremely challenging to infer.
	
	Following the findings of \citetalias{Corcho-Caballero+22}, in the present work we aim to provide a physical characterization of the different galaxy populations selected by means of the AD, and explore the connection with their evolutionary status.
	In Section~\ref{sec:obs_data} we describe the \man and \sam galaxy samples under study and the derived quantities that are used to describe their evolutionary status.
	Section~\ref{sec:m-ssfr} provides a comparison of the AD classification with the distribution of galaxies across the \mass-\ssfr plane.
	In Section~\ref{sec:cornerplot} we show the overall distribution of each AD population in terms of all the properties under study.
	Section~\ref{sec:physical_trends} presents the analysis of the trends found for each property as a function of stellar mass and environment, including the relative fraction of each AD population.
	In Sections~\ref{sec:nature_nurture} and~\ref{sec:old_or_dead} we discuss the possible dominant quenching mechanisms and the limitations of our analysis, respectively.
	Finally, we provide a summary of the main findings in Section~\ref{sec:conclusions}.
	
	Throughout this work, we adopt a $\Lambda$CMD cosmology, with $\rm H_0=70~ km\,s^{-1}\,Mpc^{-1}$ and $\Omega_m = 0.3$.
	
	\section{Methods}
	\label{sec:obs_data}
	
	\subsection{Galaxy samples}

	\subsubsection{MaNGA}
	\label{sec:MANGA}
	The Mapping Nearby Galaxies at Apache Point Observatory survey \citep[MaNGA,][]{MANGAoverview} is one of the fourth-generation Sloan Digital Sky Survey (SDSS) core programs, which was able to measure spatially resolved spectroscopy of $\sim10000$ galaxies.
	The instrument for carrying out the survey employs 17 fibre-bundle integral field units (IFU) that vary in diameter from $12''$ to $32''$ (19 to 127 fibers per IFU) with a wavelength coverage over $3600-10300$~\AA\, at $R\sim2000$ \citep{Drory+15}.
	
	The first principle that motivated the sample selection consists of getting a large enough number of galaxies to fill six bins of stellar mass, SFR and environment (216 bins) with 50 objects respectively \citep{MANGAselection, Yan+16a}.
	Additionally, galaxies are selected following a approximately flat distribution in terms of stellar mass in the range $9 < \log_{10}(M_*/M_\odot)< 12$ (based on the K-corrected $i$-band).
	In terms of the spatial extent, approximately two thirds of the total sample were observed up to \reff{1.5} (Primary sample) while $\sim$ one third is covered up to \reff{2.5} (Secondary sample).
	In addition, a third sample was selected to overpopulate the number of green valley systems; currently between the star-forming and passive populations, comprising $\sim10\%$ of the total sample.
	This considerations translate into a sample of galaxies that ranges in redshift between $0.01 \lesssim z \lesssim 0.15$.
	
	In this work we are using the complete MaNGA release, as part of the 17th SDSS data release \citep{SDSSDR17}, that comprises more than 10.000 galaxies.
	We end up with a total sample of $\sim8.600$ galaxies after selecting those with reliable \pipe estimates (see Section~\ref{sec:physical_quantities}).
	We will use the publicly available volume weights provided in \citet{Sanchez+22}, computed on an ``a posteriori'' basis as described in \citet{Rodriguez-Puebla+20}, in order to perform the statistical analysis using a volume-corrected sample in Section~\ref{sec:m-ssfr}.
	
	\subsubsection{SAMI}
	\label{sec:SAMI}
	The Sydney-Australian Astronomical Observatory Multi-object Integral field spectrograph (SAMI) galaxy survey \citep{Croom+12, Bryant+15} was carried out between 2013 to 2018 at the 3.9-m Anglo Australian Telescope at Siding Spring Observatory.
	The SAMI instrument has 13 optical fibre bundles, covering a total of $\sim 1$ degree across the FoV.
	Each bundle comprises 61 optical fibres ($1.6\arcsec$ in diameter) that cover a circular FoV of $15\arcsec$,
	feeding the AAOmega spectrograph \citep{Sharp+06}.
	The data is stored in the form of two datacubes corresponding to each of the spectrograph arms, red and blue, with a wavelength coverage of $3700-5700$ ($R\sim1800$) and $6250-7350$ \angstrom ($R\sim4300$), respectively.
	
	Roughly, two thirds of the total sample ($\sim3000$ galaxies) were selected from the GAMA equatorial fields (G09, G12 and G15), part of the Galaxy and Mass Assembly survey \citep{Driver+11}.
	Target GAMA galaxies where selected based on cuts of stellar mass ($8 \leq \logmass\leq 11$) as function of redshift ($0.004\leq z \leq 0.12$), with the Primary and Secondary (including lower stellar masses at higher redshifts) samples restricted to $z<0.095$ and $z<0.12$, respectively.
	The rest of the sample was selected from eight nearby clusters: a Primary sample  with $R<R_{200}$ and $\logmass > 9.5$ or $\logmass > 10$ for $z_{\rm clus} < 0.045$ and $z_{\rm clus} > 0.045$, respectively; and a Secondary sample  \citep[see][for details]{Owers+17}.
	In this work we use the third and final \sam data release \citep[][]{Croom+21}, comprising 2100 and 888 galaxies from the GAMA and cluster regions, respectively.
	
	\subsection{AD classification}
	
	The proxies used for computing the Ageing Diagram -- \balbreak, \ew --, were measured following the same scheme outlined in \citet{Corcho-Caballero+22}.

	The Balmer Break index at 4000\,\angstrom is defined as
	\begin{equation}
		D_n(4000) \equiv \frac{\langle F_\lambda \rangle_{4050-4250~\AA}}{\langle F_\lambda \rangle_{3850-3950~\AA}},
	\end{equation}
	where $\langle F_\lambda \rangle$ denotes the average flux density, computed within the wavelength ranges $3850-3950~\text{\AA}$, and $4050-4250~\text{\AA}$, respectively.
	
	The raw \ha equivalent width, including both absorption and emission components, is defined as
	\begin{equation}
		\label{eq:ew}
		{\rm EW(H\alpha)}
		\equiv
		\int_{\rm 6550\,\angstrom}^{\rm 6575\,\angstrom}
		\frac{ F_\lambda(\lambda) }
		{\frac{F_{\rm B}\lambda_{\rm R}-F_{\rm R}\lambda_{\rm B}}{\lambda_{\rm R}-\lambda_{\rm B}}+
			\lambda\frac{F_{\rm R}-F_{\rm B}}{\lambda_{\rm R}-\lambda_{\rm B}}}-1\ \dd\lambda,
	\end{equation}
	where $F_{\rm B}$ and $F_{\rm R}$ correspond to the mean flux per unit wavelength computed in  the $6470-6530$~\AA\ and $6600-6660$~\AA\ bands, with central wavelengths $\lambda_{\rm B}=6500$~\AA\ and $\lambda_{\rm R}=6630$~\AA, respectively. 
	Under this definition, positive and negative values of EW denote emission and absorption, respectively.
	
	To derive these quantities we use integrated elliptical aperture spectra, based on photometric S\'ersic fits, restricted to \reff{1} along the semi-major axis.
	For \sam galaxies we use the estimates provided by the collaboration \citep{Croom+21}.
	Likewise, we use the ellipticity, and effective radius reported values from the NSA catalog\footnote{\href{https://www.sdss4.org/dr17/manga/manga-target-selection/nsa/}{NASA-Sloan Atlas catalogue}} \citep{Blanton+11} to compute the integrated spectra of \man galaxies.
	
	Finally, in order to classify each galaxy in terms of the four AD domains, we employ the ageing and quenched demarcation lines presented in \citetalias{Corcho-Caballero+22}:
	\begin{eqnarray}
		\label{eq:ageing_line}
		\text{Ageing}:~ \ew / \angstrom = 250.0 \cdot 10^{-1.2 \cdot \balbreak} - 4.3, \\
		\label{eq:quenched_line}
		\text{Quenched}:~ \ew / \angstrom = -12.0 \cdot 10^{-0.5 \cdot \balbreak} + 1.8,
	\end{eqnarray}
	
	According to these expressions, galaxies located above the two lines are classified as Ageing.
	Objects placed below eq.~\eqref{eq:ageing_line} and above eq.~\eqref{eq:quenched_line} are denoted as Undetermined.
	Retired systems are placed below and above the ageing and quenched lines, respectively.
	Finally, Quenched galaxies correspond to the systems located below both demarcation lines.

	\subsection{Physical quantities}
	\label{sec:physical_quantities}

	The properties that will be used to characterise the galaxy populations derived from the AD are: the Balmber Break index, \balbreak; the raw \ha equivalent width \ew; total stellar mass \logmass;  the Petrosian radius containing the 50\% and 90\% of light in the $r$-band, respectively (\halfrad, \totrad); integrated stellar metallicity within \reff{1}, \logz; the ratio between the velocity and velocity dispersion within \reff{1}, \vsigma; fifth-nearest-neighbour surface density, \envir, and the classification between central and satellite systems.
	They comprise a set of proxies tightly connected to the current evolutionary state of each galaxy.
	Below we provide a description of the source and methodology used to derive each quantity for both samples.
	
	\subsubsection{Stellar properties}
	
	The total stellar mass, \logmass, for \man galaxies was selected from the NSA catalog \citep{Blanton+11}, estimated using multiwavelength photometry from UV to IR \citep{MANGAoverview, Sanchez+22}.
	For \sam galaxies, the stellar masses were estimated using an empirical relation between optical colours and stellar mass \citep[see eq. (3) or (6) in][for the \sam GAMA and Cluster regions, respectively]{Bryant+15, Owers+17}.
	
	We use the $r$ band Petrosian radius containing the 50 and 90 percent of total light, respectively, to characterize the spatial extent of galaxies and define the concentration index $C=\frac{R_{90}}{R_{50}}$ used as a proxy for morphological class.
	For \man galaxies we use the estimates provided in the NSA catalog.
	For \sam GAMA galaxies we employ the values provided by the GAMA collaboration\footnote{ \href{http://www.gama-survey.org/dr4/schema/table.php?id=684}{DR4 GAMA Science Catalogue}} \citep{Driver+22}.
	Regarding \sam Cluster galaxies, we only use the values of \rfifth provided in \citet{Owers+17}, as there are no publicly available estimates of \rpetro{90}.
	
	Stellar metallicities, \logz, and kinematic quantities, i.e. \vsigma, were computed by means of \pipe \citep{pypipe3d} for \man galaxies, as described in \citet{Sanchez+22}.
	For \sam, we use \ppxf \citep{ppxf} estimates of \vsigma \citep[see][]{van-de-Sande+17}, measured from luminosity-weighted aperture spectra extracted from \reff{1} circular apertures where possible, otherwise they were measured from the closest possible aperture in radius.
	We use the stellar metallicities reported in \citet{Scott+17},
	estimated within one \re  circular aperture by matching a set of observed Lick indices to various SSP models \citep{Schiavon+07, Thomas+11}.
	
	\subsubsection{Environment}
	We compute the projected density to the 5th nearest neighbour, \envir, for \man galaxies using the estimates provided by the Galaxy Environment for MaNGA Value Added Catalogue \citep[GEMA-VAC\footnote{GEMA-VAC \href{https://data.sdss.org/datamodel/files/MANGA_GEMA/GEMA_VER/GEMA.html}{Datamodel}}][]{Argudo-Fernandez+15}.
	After defining a primary volume-limited sample of galaxies, the projected distance to the fifth nearest neighbour, $d_5$, was estimated using a circular aperture of 5 Mpc and a line-of-sight velocity difference of 500 km/s.
	Since not all galaxies have a detected fifth neighbour within the adopted range (roughly $\sim30$ percent of the sample), some of the distances correspond to lower neighbour ranks (i.e. $d_5 = d_n$, with $n<5$), and their projected densities will correspond to:
	\begin{equation}
		\envir = \frac{n}{\pi d_n^2}
	\end{equation}
	
	In addition, galaxies flagged as isolated, with no density estimates, are assigned the lowest value of the sample.
	
	For \sam galaxies, we use estimates of \envir taken from \citet{Croom+21}.
	They were computed both for GAMA and Cluster regions by specifying a density-defining population with absolute $r$-band magnitudes $M_r< -18.6$ or $M_r< -19.0$ (when including the secondary targets).
	Then, the surface number density for each SAMI galaxy is determined by computing the comoving projected distances to the fifth nearest neighbour (without limiting the range to any given aperture).
	In addition, the values of \envir where corrected for spectroscopic incompleteness (with corrections smaller than $\sim15$ percent).
	
	The separation between centrals and satellites is restricted to the \sam GAMA sample and is based on the group catalog\footnote{Taken from \href{http://www.gama-survey.org/dr4/data/cat/GroupFinding/v10/}{G3CGal (v10)}} provided by \citet{Robotham+11}.
	For cluster galaxies, we will assume that all systems can be classified as satellites with the exception of the brightest central galaxies on each cluster (BCGs).

	\section{Results}
	\label{sec:results}
		\subsection{From the AD to the $M_\star-sSFR$ plane}
	\label{sec:m-ssfr}
	\begin{figure}
		\includegraphics[width=\linewidth]{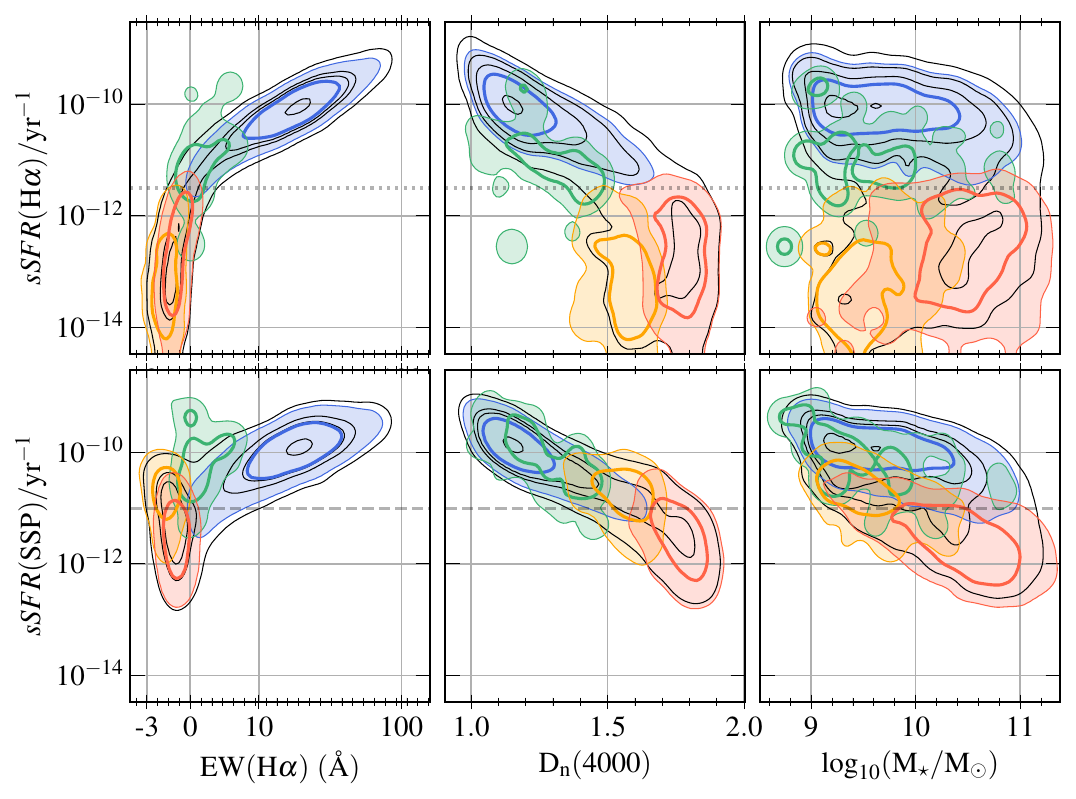}
		\caption{(left and mid columns) The relation between the AD proxies -- \ew, \balbreak -- and the integrated sSFR based on \ha, or the SSP recovered masses from \pipe for the \man sample. (right column) $M_\star-sSFR$ plane.
			The (volume-corrected) distribution of Ageing, Undetermined, Quenched and Retired galaxies are denoted as blue, green, orange and red contours (50 and 90), respectively.
			The total distribution of the sample (50 and 90 percentiles) is illustrated as black contours.}
		\label{fig:ssfr}
	\end{figure}
		\begin{figure*}
		\includegraphics[width=\linewidth]{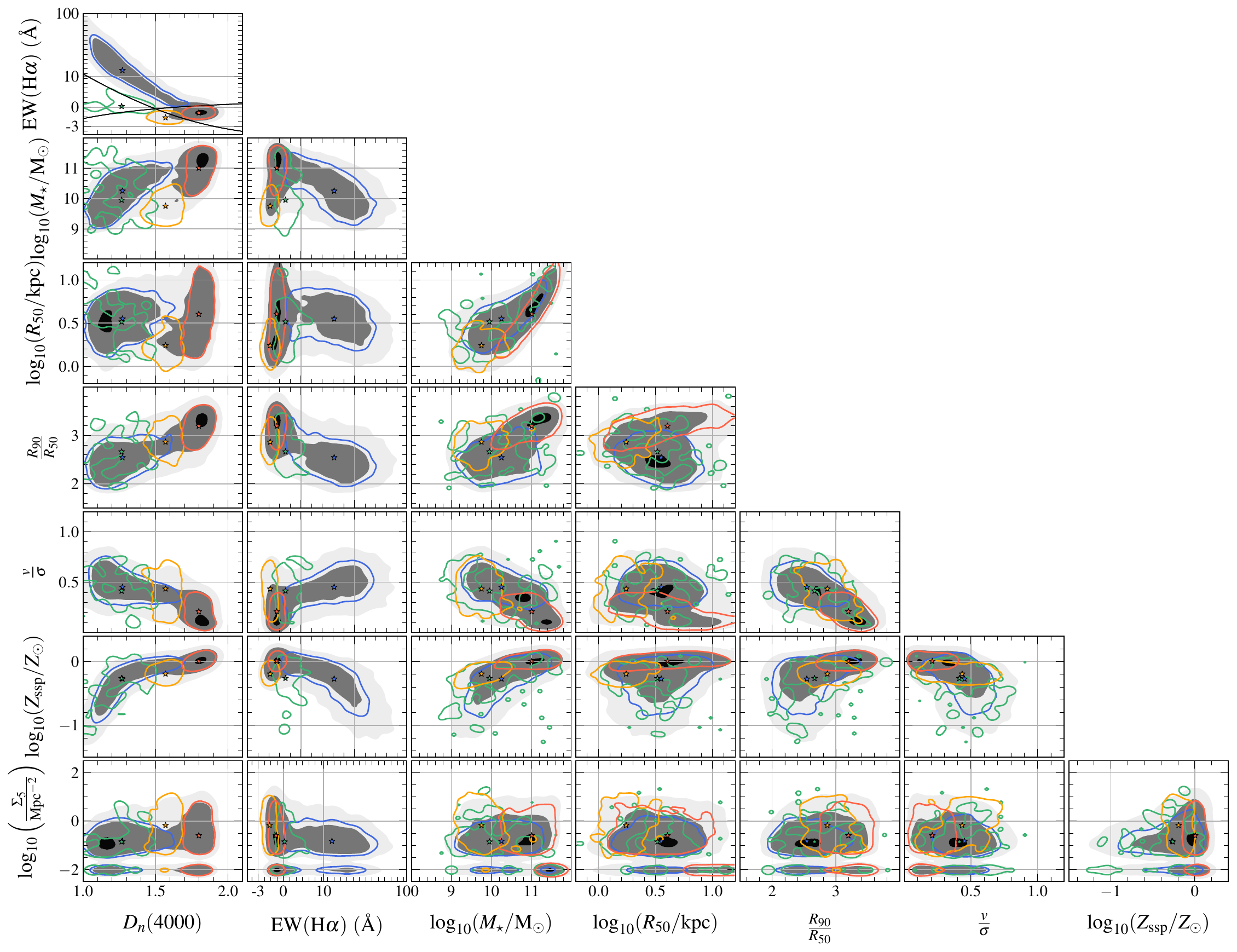}
		\caption{Distribution of physical properties of \man sample.
			The total sample distribution for each pair of properties is shown in black.
			Coloured contours denote the distribution of Ageing (blue), Undetermined (green), Quenched (orange) and Retired (red) populations, encompassing the 90 and 50 per cent of the total distribution, respectively.
			Coloured star-markers denote the median value of each distribution.
			The AD panel includes the demarcation lines included in \citetalias{Corcho-Caballero+22}.
			On the bottom-left corner of each panel the number of galaxies within each population is show using the same colour coding.}
		\label{fig:manga_cornerplot}
	\end{figure*}
	
	\begin{figure*}
		\includegraphics[width=\linewidth]{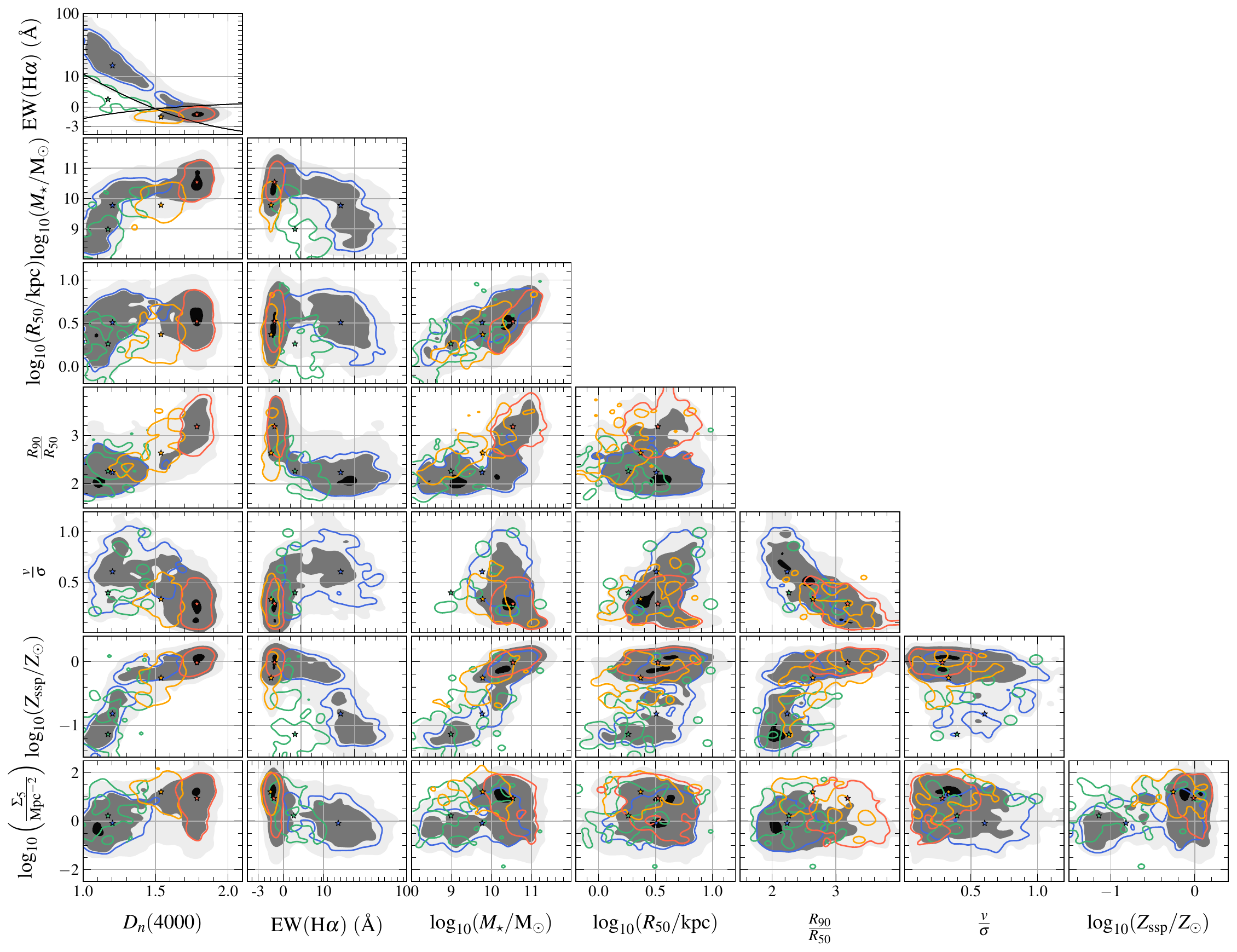}
		\caption{Same as Figure~\ref{fig:manga_cornerplot} for \sam sample.
		}
		\label{fig:sami_cornerplot}
	\end{figure*}
	
	In this section we show the connection between the AD classification and the $M_\star-sSFR$ plane.
	We restrict our analysis to the \man sample as we apply the volume corrections provided in \citet{Sanchez+22} in order to compare with previous work \citep[e.g.][]{Corcho-Caballero+20}.
	We employ two estimates of the $sSFR$ that correspond to different time averages of the recent star formation history of the galaxy.
	On the one hand, we use the $sSFR$ derived from the intensity of the \ha line, tracing in principle the fraction of O and B stars that ionise the ISM.
	This roughly corresponds to an average of the SFR over the last $\lesssim20-30$ Myr \citep{Kennicutt98, Davies+16}; $sSFR(\ha)$.
	On the other hand, we use SFR estimates derived from the stellar population synthesis using \pipe, that accounts for the average SFR over the last $\sim 30-100$ Myr; $sSFR(\rm SSP)$.
	More precisely, we use the logarithmic mean between these two values, where we observe the best match with  $sSFR(\ha)$ for AGs.
	
	As expected, we show in Figure~\ref{fig:ssfr} that both of our observational proxies present a strong correlation with the $sSFR$ estimates.
	By construction, \ew is directly connected to $sSFR(\ha)$, and galaxies with $\ew > 0$ show a one-to-one correspondence between both quantities.
	Below this threshold, roughly equivalent to $sSFR\sim 3\times 10^{-12}~\rm yr^{-1}$, the estimates of $sSFR(\ha)$ are dominated by the residual emission after \pipe subtracts the stellar continuum.
	These measurements are extremely uncertain, and it is not possible to discriminate Quenched and Retired systems, or even completely dead galaxies \citep{Corcho-Caballero+21a}.
	However, when using $sSFR(\rm SSP)$ -- tightly correlated with \balbreak -- we find that the majority of Quenched and Retired galaxies display values above and below $\sim 10^{-11}~\rm yr^{-1}$, respectively, albeit the separation according to this threshold, often used in the literature to discriminate between ``star-forming'' and ``passive'' populations, is far from perfect: $sSFR(\ha) < 3\times 10^{-12}~\rm yr^{-1}$ selects both Quenched and Retired systems, whereas $sSFR(\rm SSP) < 10^{-11}~\rm yr^{-1}$ includes a mixture of Ageing and Retired objects.
	
	In the $M_\star-sSFR$ plane, Ageing galaxies arrange along the MS, with mean values around $10^{-10}~\rm yr^{-1}$.
	Undetermined systems are on average below the Ageing population, but they would still be classified as MS systems, while Quenched and Retired galaxies correspond to a population that extends from the so-called Green Valley (i.e. $sSFR \sim 10^{-11}~\rm yr^{-1}$) to residual levels of star formation.
	However, the overall bi-dimensional distribution across the plane is strongly dependent on the SFR tracer and the corresponding timescale \change{\citep[see e.g.][]{KennicuttEvans12, Donnari+19, Caplar+19}}.
	In our case, $sSFR(\ha)$ is obviously more appropriate than $sSFR(\rm SSP)$ to identify QGs, but imposing a threshold based on only one timescale (even if the adopted threshold depends on $M_\star$, parallel to the MS) is not guaranteed to identify galaxies that suffered some quenching event, as it will include a fraction of Retired galaxies that may also form through the Ageing channel.
	Moreover, we would also like to note that measurement errors, as well as physical fluctuations in the star formation rate \citep[e.g.][]{Davies+19, Davies+22}, may play a role on the observed sSFR on short timescales.

	\subsection{Physical properties in terms of AD class}
	\label{sec:cornerplot}

	The main purpose of this work is to provide a physical characterization of the galaxy populations classified by the Ageing Diagram (AD), following the scheme presented in \citet{Corcho-Caballero+22}.
	We have selected a range of physical properties that are known to be tightly connected with the evolutionary status of galaxies (see \ref{sec:physical_quantities}).
	In order to provide a summarized view, we show in Figures~\ref{fig:manga_cornerplot}~and~\ref{fig:sami_cornerplot} the distribution of each pair of quantities for \man and \sam galaxy samples, respectively.
	\change{Both samples are kept separated during the rest of the paper.
	This is done to prevent potential biases driven by their distinct sample selection criteria, as well as the different methodological approaches employed to derive the physical quantities.
	}
	The total probability distribution is illustrated by the grey solid contours, while the individual distributions of Ageing, Undetermined, Quenched and Retired populations are denoted as coloured contours.
	The galaxy classification is based on their location across the plane formed by the \balbreak-\ew.
	This is most clearly seen on the top left panels of both figures, where we include the demarcation lines (black solid lines) used to classify the galaxies into four different domains as presented in \citetalias{Corcho-Caballero+22}.
	
	As a general remark, we find that both samples provide broadly consistent distributions at all panels.
	Table~\ref{tab:percentiles} also includes the median and dispersion values (16 and 84 percentiles) for each property, in terms of Ageing, Undetermined, Quenched and Retired populations, in both observational samples.
	
	As shown before, the distribution as function of total stellar mass, \logmass, and \balbreak or \ew is tightly connected to the distribution of galaxies in terms of \logmass and $sSFR$.
	The majority of previous studies focusing on the $M_\star-sSFR$ relation \citep[e.g.][]{Peng+10} usually divided the sample into star-forming ($sSFR\gtrsim 10^{-11}~\rm yr^{-1}$) and passive populations ($sSFR\lesssim 10^{-11}~\rm yr^{-1}$), which could be interpreted as the division between galaxies with values above or below $\balbreak\sim1.6$ or $\ew\sim0$, in line with the bimodal paradigm where galaxies are either hosting star formation processes or not \citep[see e.g.][]{Corcho-Caballero+21a}.
	Although in statistical terms this seems a reasonable approach, as the overall probability distributions of both \balbreak and \ew, integrated over all galaxy masses, show two distinct peaks, here we show that our diagram is able to go one step further in discriminating between those systems that undergone recent quenching (i.e. a sharp truncation of the SFH) and those that evolved secularly over the last $\sim3$Gyr \citep[with smoothly declining SFHs, e.g.][]{Abramson+16}.
	
	Under our prescription, the Ageing domain roughly corresponds to the so-called Main Sequence of Star Forming galaxies (MS), extending from low to high mass systems ($\mass \le 10^{11}~\Msun$) whose star formation activity is consistently traced by our two observational proxies, probing different time scales.
	On the other hand, as mentioned above, the Quenched population is predominantly made of galaxies in the mass range $9 \lesssim \logmass \lesssim 10.5$, forming stars according to \balbreak but featuring negligible $sSFR$ in terms of \ew.
	Both populations, clearly separated in the AD, converge to the Retired domain towards the high mass end, thus implying that not all red sequence systems were necessarily quenched in the past.
	The Retired domain  -- mainly composed of massive galaxies with $\logmass \gtrsim 10$ -- harbours a combination of formerly quenched systems and \change{formerly Ageing systems that have gradually consumed most of their gas reservoir}.
	
	Regarding the spatial extent of stars in galaxies, we find that Quenched systems have systematically smaller \rfifth -- roughly between 1 to 3 kpc -- than the rest of the populations, that span over a wide range of values from 1 to 10 kpc.
	On the other hand, the plane defined by the concentration index \concen and \rfifth presents a ``V'' shape.
	Ageing and Retired galaxies roughly correspond to the edges (illustrating the classical distinction between disk and spheroidal morphologies), while QGs clump around the vertex (small and compact objects).
	In terms of their kinematic morphology, Quenched galaxies are also found to display intermediate values of \vsigma compared to the Ageing ($\vsigma \gtrsim 0.3$) and Retired populations ($\vsigma \lesssim 0.3$).
	In the \man sample, QGs present closer values of \vsigma to the Ageing pop. (i.e. more rotationally supported), while the opposite is true for \sam galaxies.
	
	The average metallicity of the stellar population, \logz, also appears to be connected with the distribution of galaxies along the AD.
	For both \man and \sam samples we find that the Ageing domain hosts more chemically primitive systems, while Retired galaxies present higher metallicities close to solar values \citep[in agreement with previous results, e.g.][]{Ascasibar+15, Corcho-Caballero+20}.
	Systems classified as Quenched arrange above Ageing galaxies in the \logmass vs \logz plane, corresponding to a (low-mass) chemically evolved population.
	
	Finally, we find that the environment, characterized by \envir, also yields a strong correlation with the AD classification scheme.
	Although both observational samples target different environments, they show a consistent result: Quenched galaxies are primarily found in denser environments than the Ageing or Retired populations.
	
	\subsection{Physical trends and fractions}
	\label{sec:physical_trends}
	
	In this section we explore in detail the relation between stellar mass, interpreted as the fundamental tracer of the evolutionary state of a galaxy, with the rest of physical properties under study in this work: \balbreak, \rpetro{50}, \concen, \vsigma, \envir.
	In addition, we further study the impact of environment by binning \man and \sam samples into centrals and satellites (see \ref{sec:physical_quantities} for details).
	
	Figures \ref{fig:binned_centrals_d4000} to \ref{fig:binned_centrals_logsigma5} illustrate the relative fraction of galaxies classified within each AD domain as function of each pair of properties.
	For illustrative purposes, since the contribution of Undetermined galaxies is negligible ($<1$ percent), we will only consider the Ageing, Quenched, and Retired populations in this analysis.
	Complementarily to the domain fraction, the trends with total stellar mass along each diagram are estimated using the running percentiles (16, 50, 84) illustrated by black dashed lines.
	Finally, we include contours that comprise 50 and 90 percent of the sample at every panel (black solid lines), tracing the bulk of the distribution (that can also be used to asses the confidence limits of the running percentiles).
	
	\subsubsection{\logmass and \balbreak}
	
\begin{figure}
		\includegraphics[width=\linewidth]{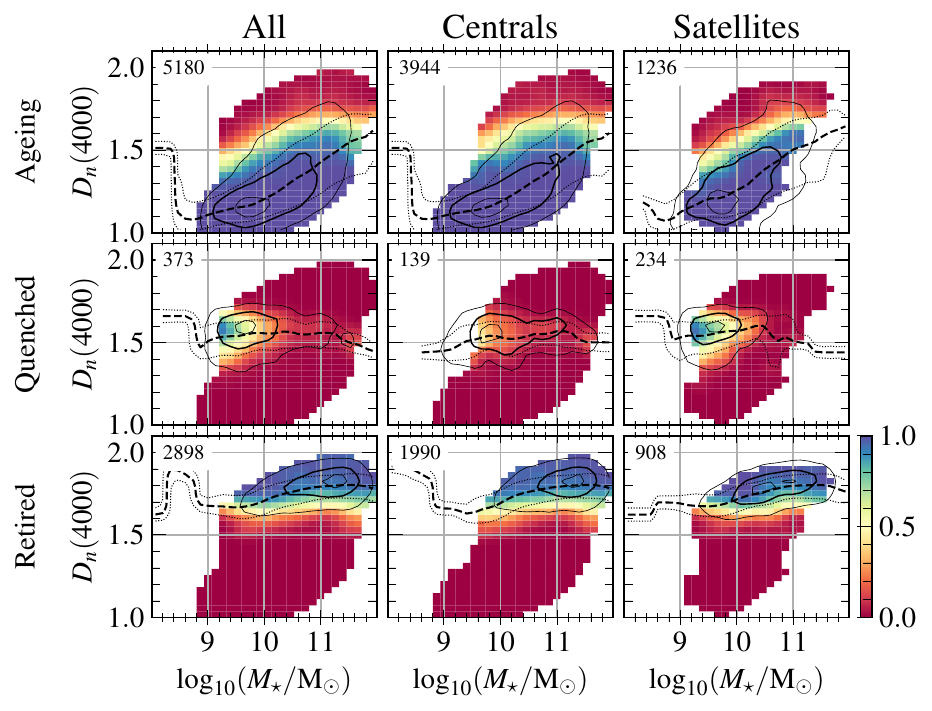}
		\includegraphics[width=\linewidth]{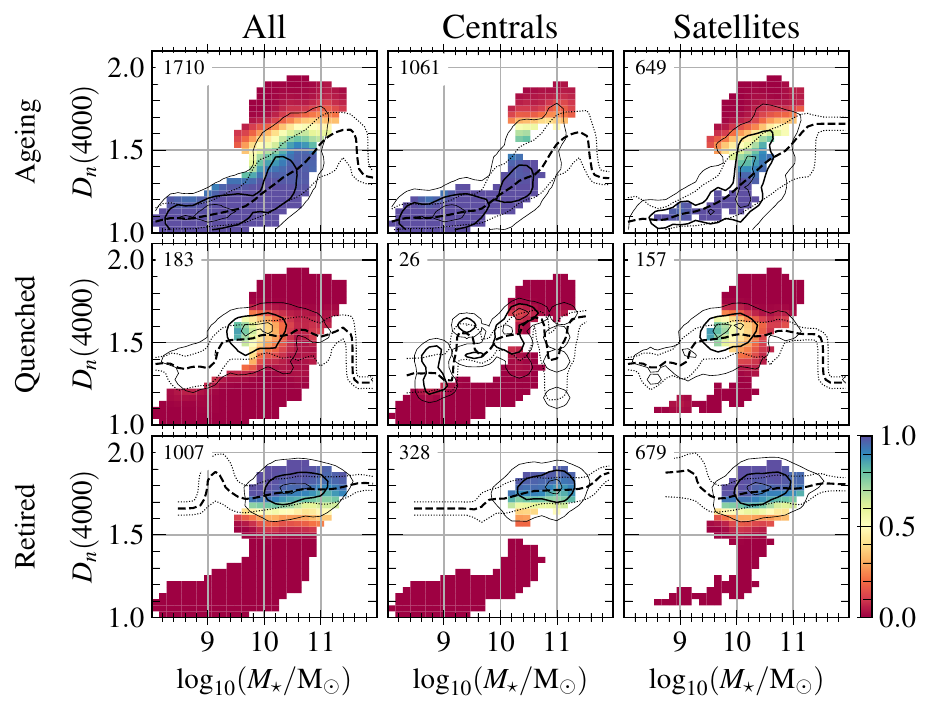}
		\caption{Fraction of Ageing, Quenched and Retired galaxies as function of \logmass and \balbreak for \man (top) and \sam (bottom) samples. 
		The first column includes all galaxies, while the rest split the sample into Isolated, Centrals and Satellites bins.
		Coloured maps show the relative fraction of each galaxy class, \change{smoothed with a Gaussian kernel (standard deviation equal to 1/30 of the dynamical range of the axes), for those bins including a total of more than 3 objects}.
		Solid black lines encompass 90 and 50 percent of the galaxies on each panel (the total number is included on the top-left corner).
		Dashed and dotted lines represent the running median and 16/84 percentiles as function of \logmass.
        \change{Most Quenched galaxies are low-mass satellites in the Green Valley (which is nonetheless dominated by the Ageing population). Quenched centrals span the whole mass range.}
		}
		\label{fig:binned_centrals_d4000}
	\end{figure}

	As mentioned in the previous section, the connection between stellar mass and \balbreak roughly corresponds to the $M_\star-sSFR$ relation.
	Figure~\ref{fig:binned_centrals_d4000} shows a good agreement between \man and \sam samples, although their mass distributions are slightly different.
	
	Ageing galaxies distribute along a wide range of masses and values of \balbreak, showing increasing values of \balbreak (lower $sSFR$) as mass increases \citep[akin to the distributions of $sSFR$ shown in][]{Eales+18a, Corcho-Caballero+20}.
	The population is dominated by, but not limited to, low-mass galaxies reflecting the so-called downsizing effect.
	By construction, the fraction of Retired galaxies below $\balbreak\sim1.6$ is null, and most Quenched galaxies cluster around that value.
	
	On the other hand, we also note that the vast majority of Quenched galaxies correspond to Satellite systems, albeit a non-negligible fraction of this population (37 and 14 per cent for \man and \sam, respectively) have been classified as Central objects.
	It is interesting to note that most Quenched satellite galaxies are typically below $\sim 10^{10}$~\Msun, whereas centrals span the whole mass range of the sample, even beyond $10^{11}$~\Msun.
	
	\subsubsection{Morphology}
	
	\begin{figure}
		\includegraphics[width=\linewidth]{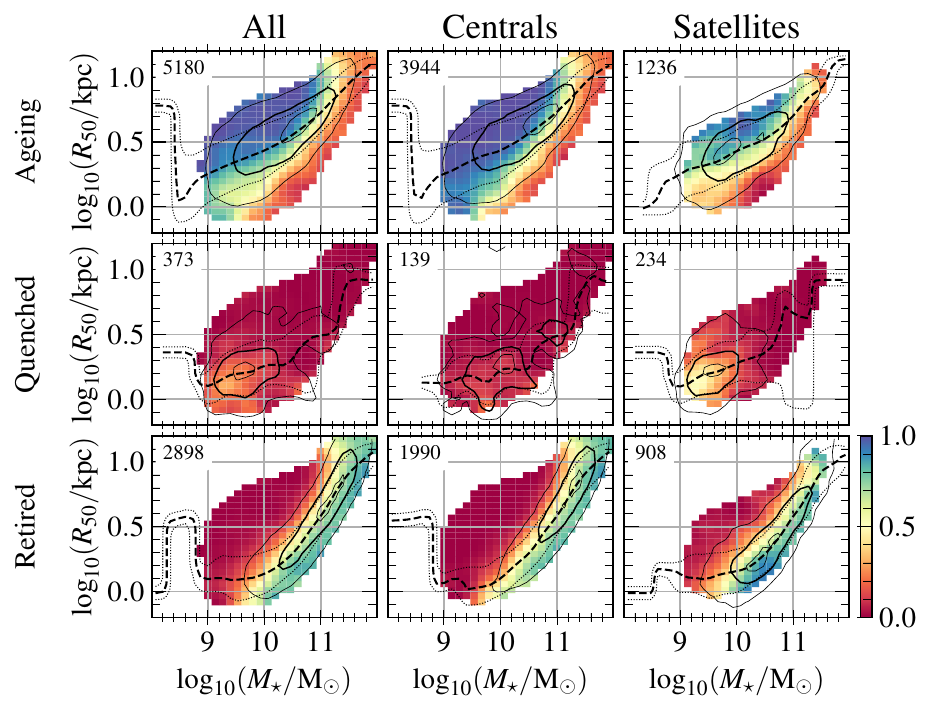}
		\includegraphics[width=\linewidth]{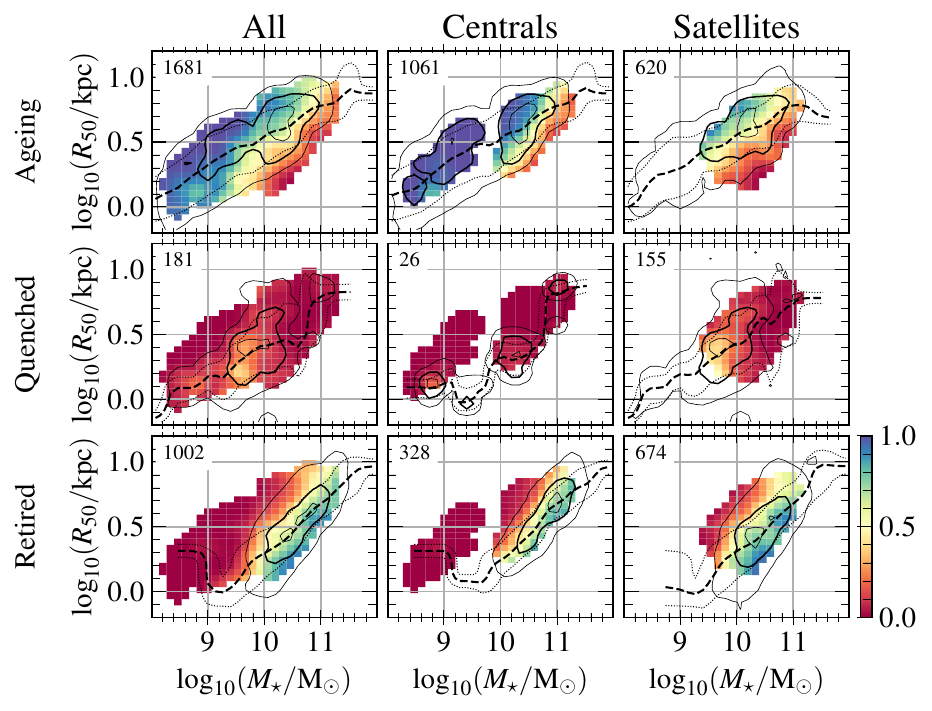}
		\caption{Fraction of Ageing, Quenched and Retired galaxies as function of \logmass and \logre for \man (top) and \sam (bottom) samples (see Fig.~\ref{fig:binned_centrals_d4000} for more details).
        \change{Quenched galaxies are more compact than Ageing systems. Retired galaxies feature a surface density $\sim 200-300$~\Msun~pc$^{-2}$.}
        }
		\label{fig:binned_centrals_logre}
	\end{figure}
	
	\begin{figure}
		\includegraphics[width=\linewidth]{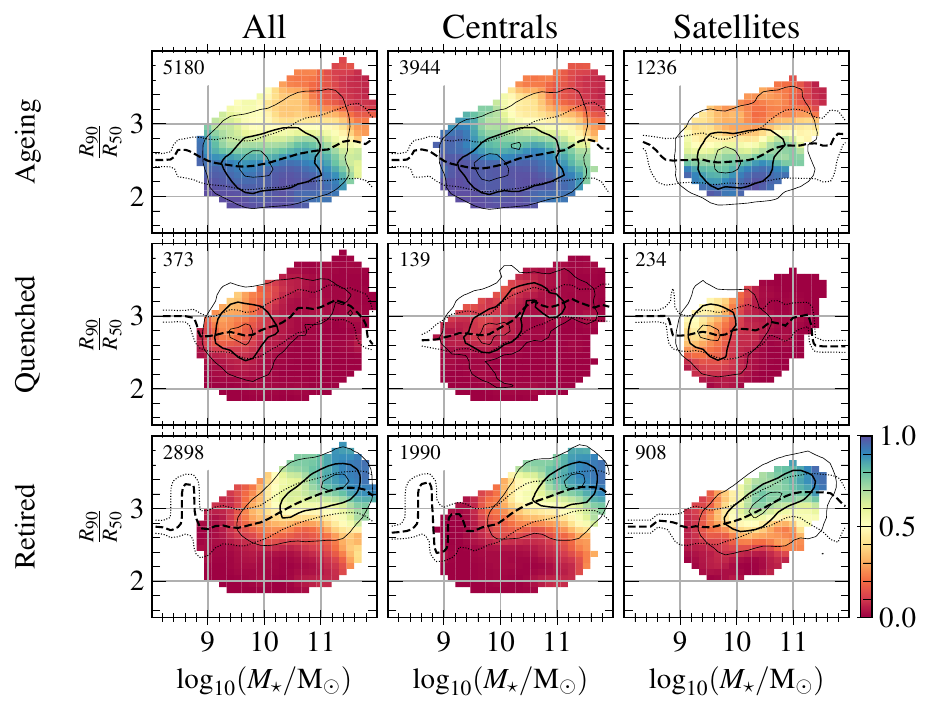}
		\includegraphics[width=\linewidth]{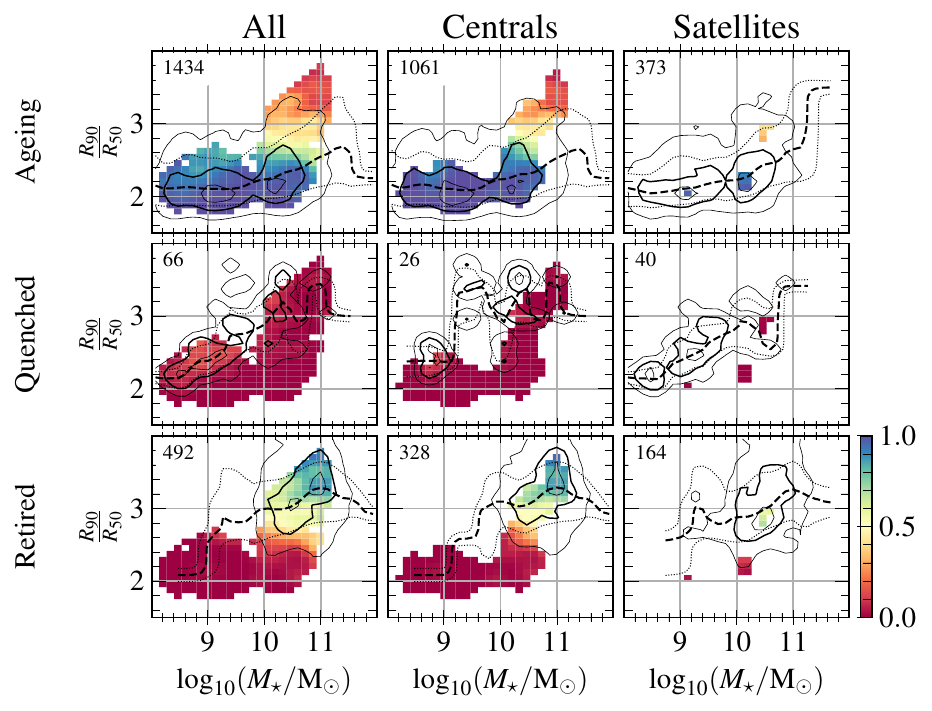}
		\caption{Fraction of Ageing, Quenched and Retired galaxies as function of \logmass and \concen for \man (top) and \sam (bottom) samples (see Fig.~\ref{fig:binned_centrals_d4000} for more details).
        \change{Ageing systems feature $\concen\sim 2.2-2.5$ at all masses. Quenched and Retired galaxies are more concentrated, and they share a positive correlation bewtween $M_\star$ and \concen.}
        }
		\label{fig:binned_centrals_concentration}
	\end{figure}
	The spatial distribution of stars within galaxies, traced by their characteristic radius (i.e. \rpetro{50}) and light concentration (i.e. \concen) is another fundamental aspect connected to their assembly history.
	
	We show in Figure~\ref{fig:binned_centrals_logre} the plane defined by the total stellar mass and effective radius, often referred to as the mass-size relation.
	There is a sharp transition between the fraction of Ageing systems in favour of the Retired population \change{(note the abrupt change in the colour maps)}, roughly corresponding to a stellar surface density threshold $\Sigma_{\rm Q} \equiv \frac{M_\star/2}{\pi \rfifth^2} \sim 200-300~\rm \Msun~pc^{-2}$.
	While Retired galaxies arrange along a very tight relation, $\rfifth \propto M_\star^{0.5}$, roughly coincident with such threshold, Ageing systems present a much larger scatter, illustrated by the running percentiles, consistent with a rich diversity of disk-like morphologies \citep[see e.g.][]{Shen+03, van-der-Wel+14}.
	
	As mentioned before, the majority of Quenched systems correspond to compact, low-mass, satellite galaxies.
	Here we see that their distribution in the mass-size plane is located at an intermediate region between the Ageing and Retired populations; i.e. $9 < \logmass < 10$ and $0 < \logre < 0.5$.
	Central QGs, on the other hand, seem to be located close to the characteristic stellar mass surface density $\Sigma_{\rm Q}$ over the whole mass range \change{(see running-median values traced by the dashed lines)}.
	
	When exploring the concentration of light \concen, plotted in Figure~\ref{fig:binned_centrals_concentration}, we find an almost flat trend with \logmass for Ageing galaxies, with the median roughly located at $\concen \simeq 2.5$ and $\concen \simeq 2.2$ for \man and \sam, respectively.
	On the other hand, the concentration of Retired systems shows a mild positive correlation with stellar mass, with the bulk of the population presenting values $\concen \gtrsim 3$.
	In between, Quenched galaxies scatter roughly around $\concen\sim 3$ and seem to be well described by the Retired mass-concentration relation both for central and satellite galaxies, \change{i.e. a mild positive trend between \concen and \logmass}.
	Note that the value of \rpetro{90} is not available for \sam cluster galaxies, and therefore the number of galaxies in this plot (especially on the \change{SAMI} panels) is significantly lower than in the other figures.
	
\begin{figure}
		\includegraphics[width=\linewidth]{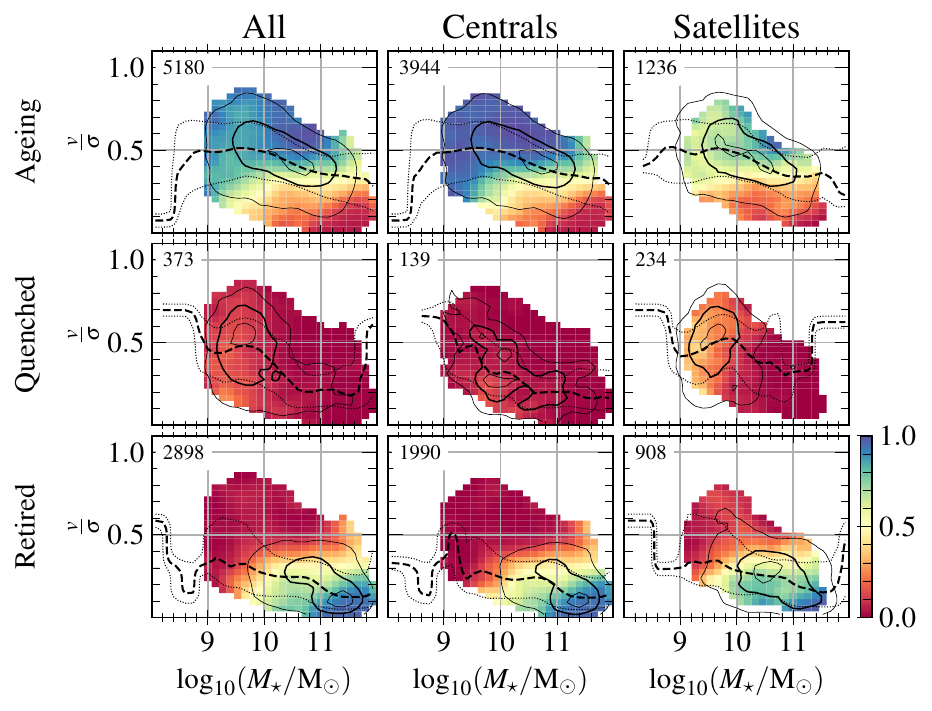}
		\includegraphics[width=\linewidth]{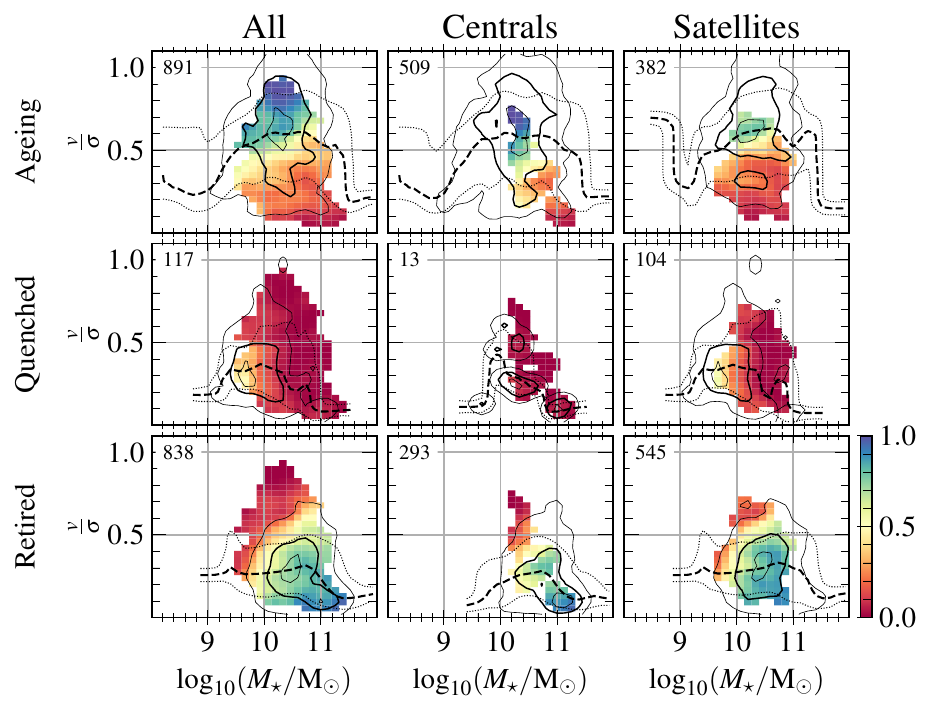}
		\caption{Fraction of Ageing, Quenched and Retired galaxies as function of \logmass and \vsigma for \man (top) and \sam (bottom) samples (see Fig.~\ref{fig:binned_centrals_d4000} for more details).
        \change{Ageing galaxies are more rotationally-supported than the Quenched and Retired population.}
        }
		\label{fig:binned_centrals_vsigma}
	\end{figure}

	Regarding the ratio between ordered to random motions within galaxies, traced by \vsigma and shown in Figure~\ref{fig:binned_centrals_vsigma},
	Ageing systems seem to display an overall correlation, most evident in \man, in the sense that more massive galaxies tend to be less rotationally supported.
	However, there is significant scatter along this trend, and one can find AGs with very different values of \vsigma.
	In contrast, Retired galaxies are almost invariably dominated by random motions, with $\vsigma < 0.5$.
	
	Quenched galaxies feature a more complex behaviour.
	While we do not find any significant difference between the mass-size and mass-concentration relations of central and satellite QGs (aside from the total number of galaxies within each class), here we see that central Quenched systems tend to be more pressure-supported than Ageing galaxies \change{(see median values of each galaxy domain)}, displaying values of \vsigma compatible with the distribution of the Retired population.
	On the other hand, the dynamical state of satellite QGs appears to be closer to AGs, \change{as they present slightly larger values of \vsigma}, hinting that the physical mechanism responsible for quenching may be different for central and satellite objects.

	\subsubsection{Chemical composition}
	
\begin{figure}
		\includegraphics[width=\linewidth]{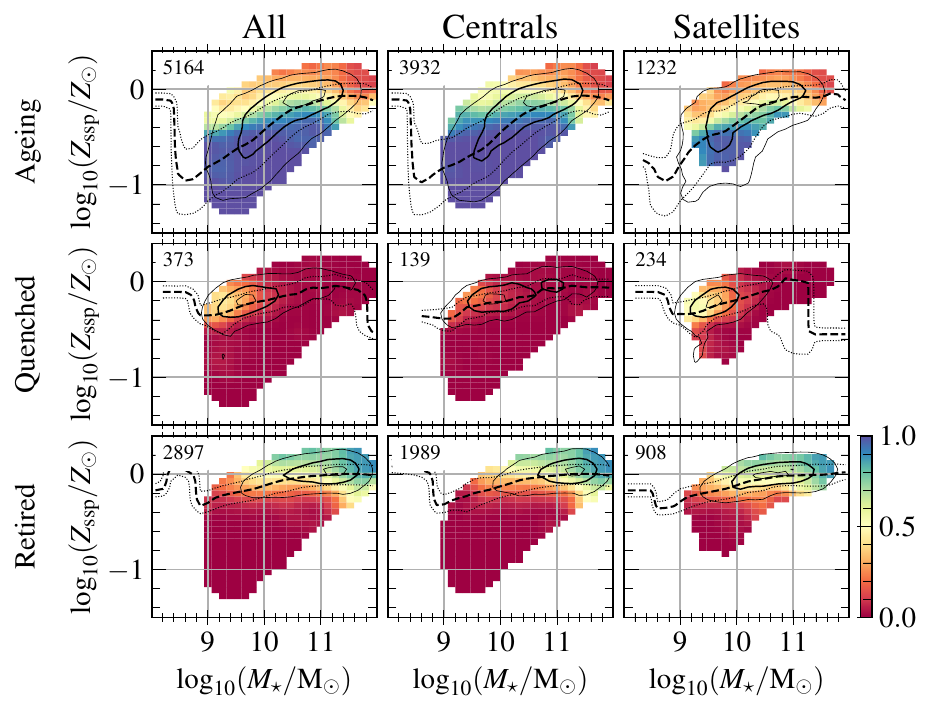}
		\includegraphics[width=\linewidth]{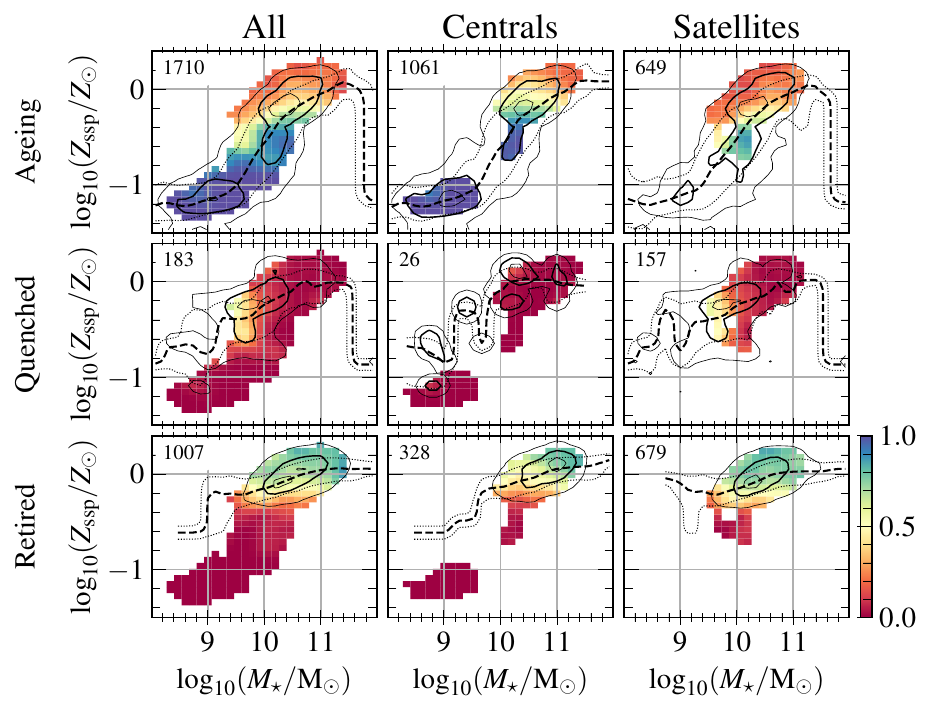}
		\caption{Fraction of Ageing, Quenched and Retired galaxies as function of \logmass and \logz for \man (top) and \sam (bottom) samples (see Fig.~\ref{fig:binned_centrals_d4000} for more details).
        \change{The mass-metallicity relation of Quenched and Retired galaxies has higher normalisation and flatter slope than the Ageing counterpart.}
        }
		\label{fig:binned_centrals_logz}
	\end{figure}

	The relation between total stellar mass and mass-weighted metallicity of the stellar population is shown in Figure~\ref{fig:binned_centrals_logz}.
	For Ageing galaxies, both samples follow the well-known trend that more massive systems are more chemically evolved, albeit \sam seems to feature a steeper slope than \man, possibly due to differences in the analysis approach as well as sample selection.
	RGs, on the other hand, concentrate at high metallicities $\logz \gtrsim -0.3$, with a much shallower slope.
	From our data, it is difficult to conclude whether Quenched galaxies represent an intermediate population between AG and RG, or they follow the same tight $\mass-\logz$ relation as the latter.
	\change{What becomes clear is that, for any given stellar mass, QGs tend to be more metal-rich than their Ageing counterparts, as illustrated by their respective running median values (black dashed lines), in agreement with previous results \citep[e.g.][]{Trussler+20}.}
	
	Once again, we do not find any significant difference between centrals and satellites regarding these chemical evolution trends.
	
	\subsubsection{Environment}
	
\begin{figure}
		\includegraphics[width=\linewidth]{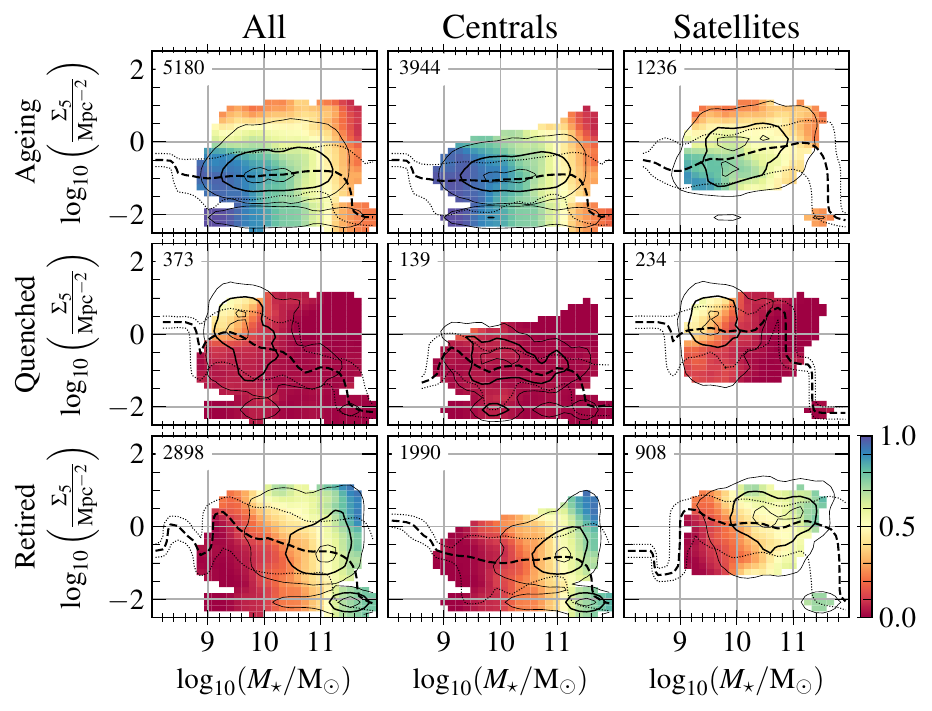}
		\includegraphics[width=\linewidth]{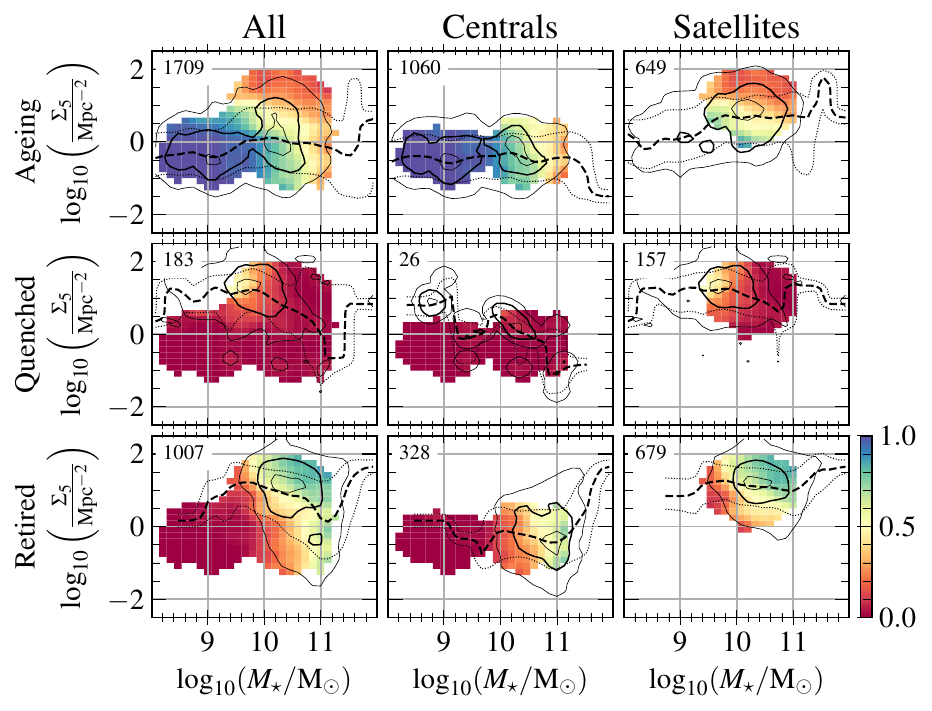}
		\caption{Fraction of Ageing, Quenched and Retired galaxies as function of \logmass and \logenvir for \man (top) and \sam (bottom) samples (see Fig.~\ref{fig:binned_centrals_d4000} for more details).
        \change{Retired galaxies dominated at high masses and/or dense environments. Quenched centrals (of any mass) are predominantly found in low-density environments.}
        }
		\label{fig:binned_centrals_logsigma5}
	\end{figure}

	In Figure~\ref{fig:binned_centrals_logsigma5}, we show the relation between \logmass and the environment density, traced by the surface density estimated from the distance to the fifth nearest neighbour, \envir.
	As expected, the majority of satellite galaxies are found in denser environments, while the opposite is true for centrals.
	When comparing the distributions of each AD population, both \man and \sam samples roughly cover the same region of the parameter space, as shown by the contour that contains the 90 percent of the distribution (thin solid line).
	However, the bulk of the distribution for \sam is located at denser environments at all panels, as roughly one third of the sample lives in clusters.
	
	\change{As illustrated by the colour maps, representing the fraction of galaxies in each class}, Ageing galaxies dominate the low- mass ($\mass < 10^{11}$~\Msun) and density ($\envir < 1~\rm Mpc^{-2}$) region of the diagram.
	Retired galaxies dominate at dense environments and high stellar masses.
	Their fraction, plotted on the bottom row of each panel, is consistent with the results provided in Fig. 6 of \citet{Peng+10}.
	
	Here we provide a more detailed characterization by including the population of recently quenched galaxies.
	As can be seen in the middle row of \man and \sam panels, we find that Quenched systems dominate the low-mass and high-density region of the diagram.
	In addition to this population, mainly composed of satellite galaxies, the large statistics in the \man survey makes it possible to uncover a non-negligible number of central QGs, spanning the whole mass range from $\logmass\sim 9$ to 11, that reside in the field ($\envir < 1~\rm Mpc^{-2}$).

	\section{Discussion}
	\label{sec:discussion}
	
	\subsection{Quenched galaxies: nature or nurture?}
	\label{sec:nature_nurture}
	
	While the correlations between the physical properties of QGs reported in this study, compared to those of Ageing and Retired systems, offer some insight on the mechanism(s) that are ultimately responsible for quenching star formation in a formerly Ageing galaxy, inferring the actual causal relations behind them is far from straightforward.
	
	Nevertheless, we would like to claim that the results shown in Section~\ref{sec:results} suggest that quenching is an ubiquitous process in the local Universe, but rare, although the precise number of galaxies that transit through this evolutionary path remains uncertain \citepalias[between 3 to 10 per cent at $z\leqslant 0.1$ according to][]{Corcho-Caballero+22}.
	In particular, we do find QGs both at high and low stellar masses, as well as galaxy density environments.
	We interpret this fact as evidence for at least two different, independent quenching mechanisms, broadly consistent with the classification proposed by \citet{Peng+10}.
	
	On the one hand, we found that the majority of Quenched galaxies are low-mass systems in dense environments.
	In order to bring AGs to the Quenched domain, this processes must act on relatively short timescales ($\lesssim 1$ Gyr).
	\change{Previous works, such as \citet{Wetzel+13} suggested that satellite systems undergo a ``delayed-then-rapid'' quenching phase consisting of an early-infall phase (2-4 Gyr) where star formation proceeds roughly unaffected, followed by a fast truncation of the SF.
	The results presented in this work are compatible with this scenario as far as the satellite population is concerned.}
	This suggests that environmental effects such as ram pressure stripping might be the main cause of the cessation of star formation in the local Universe.
	Other environmental processes, such as strangulation, would take several Gyr, and the affected galaxies would evolve from the Ageing to the Retired domain in the AD.
	As discussed in \citetalias{Corcho-Caballero+22}, QGs have spent, on average, more time in a ``starburst'' evolutionary stage (birth rate parameter $b \equiv t sSFR(t) > 2$) than AGs of the same colour.
	Here we see that QGs are also more compact, concentrated, and metal-rich than AGs of the same mass.
	All these results are consistent with a post-starburst scenario, which provides further support to the ram-pressure hypothesis.
	
	On the other hand, we also find a population of Quenched central galaxies that tend to be found in the field (37 and 14 per cent of \man and \sam QGs, respectively).
	Environmental processes are thus very unlikely to explain the sudden decline of the SFR observed in these objects.
	These galaxies span the whole mass range of our sample, and to some extent they are consistent with the simple ``mass quenching'' scenario envisaged by \citet{Peng+10}, where the quenching probability per unit time is proportional to the star formation rate.
	
	However, the mass-size relation of QGs is not trivial to justify under this hypothesis.
	An alternative explanation, along the lines of ``morphological quenching'' \change{\citep{Martig+09, Gensior+20}}, would be that star formation ceases upon a critical mass surface density $\Sigma_{\rm Q}$ is reached.
	This mechanism would equally apply to centrals and satellites, whose mass-size relation would simply trace this critical value, as we observe in Figure~\ref{fig:binned_centrals_logre}.
	One interesting question under this interpretation would be the role of environment.
	More precisely, whether galaxies in groups tend to be more compact due to externally-induced processes (nurture), or they simply evolved on shorter timescales over cosmic history because they arise from a higher overdensity peak in the primordial Universe (nature).

	\subsection{Retired galaxies: old or dead?}
	\label{sec:old_or_dead}
	
	In principle, there are two different evolutionary paths that would take a galaxy to the Retired domain: on the one hand, star formation could slowly decrease, without any significant event throughout its life (Ageing), or the SFR could suddenly drop due to a well-defined, discrete event (Quenching).
	The AD, by itself, is not sufficiently sensitive to discriminate between both scenarios, as stellar populations older than $\gtrsim 1-3$~Gyr display nearly identical optical features.
	One way to tackle this issue would be to investigate the AD at different redshifts and trace the evolution of galaxies back in time, taking advantage of current and forthcoming optical spectroscopic surveys \citep[e.g.][]{Lilly+07, LeFevre+13, Davies+18}.
	
	A related question, still poorly understood, is whether ageing and/or quenching processes are able to fully halt the conversion of gas into stars or, in contrast, Retired galaxies eventually reach some asymptotic regime of residual star formation \citep[e.g.][]{Salvador-Rusinol+20}.
	As show in \citetalias{Corcho-Caballero+21a}, many state-of-the-art cosmological simulations predict that the majority of systems off the MS have sSFR values below the resolution limit (compatible with $sSFR=0$), while observational estimates (arguably upper limits) suggest that galaxies may retain some star formation activity.
	In order to address this question, alternative tracers of star formation must be sought.
	
	\section{Summary and conclusions}
	\label{sec:conclusions}
	

	Following the classification scheme presented in \citet{Corcho-Caballero+22}, in this work we study the physical properties of the galaxy populations found in different domains of the Ageing Diagram (AD), using \man and \sam surveys.
	The AD is based on the combination of two star formation proxies, sensitive to different timescales, such as the raw \ew, tracing the \ssfr on short timescales ($\sim$ Myr), and the \balbreak, that provides an estimate of the \ssfr on $\sim$ Gyr scales.
	This method allows to detect abrupt changes on the recent star formation history of galaxies and investigate quenching episodes during the last $\sim 3$ Gyr.
	
	We use the empirical lines defined in \citetalias{Corcho-Caballero+22} to classify galaxies as Ageing (AGs), Undetermined (UGs), Quenched (QGs) or Retired (RGs).
	For each population, we explore the trends with total stellar mass, 50 per cent $r$-band light radius \rfifth, concentration \concen, ratio between ordered to random motions \vsigma, stellar chemical composition \logz, and environment, traced by the projected number density \envir and group rank (satellite vs central).
	The characteristic properties of AGs, RGs and QGs are summarized below:
	
	Ageing galaxies, whose evolution is mainly driven by secular processes, extend over the whole range of stellar mass ($10^{8.5}\leqslant \mass/\Msun \leqslant 10^{10}$) and \balbreak.
	Most AGs are classified as centrals living in the field, but we find a relatively large number of satellites as well.
	The probability distributions in terms of stellar mass and size, concentration or \vsigma, respectively, are consistent with having predominantly late-type morphologies.
	They display the well-known correlations between stellar mass, star formation and metallicity characteristic of MS galaxies.
	
	Retired objects are chemically evolved, featuring high values of \balbreak and \logz, with a very mild dependence, if any, on stellar mass.
	They arrange along a tight sequence in the mass-size plane, roughly given by $\rfifth \propto \mass^{0.5}$, at the compact end of the AGs; both populations are indeed well separated by a stellar surface density threshold $\Sigma_{\rm Q} \sim 200-300~\Msun~\rm pc^{-2}$.
	RGs are consistent with being composed by early-type systems, and they are found in all kinds of environments.
	
	The Quenched galaxy population is mainly composed by compact low-mass satellite galaxies, \change{in good agreement with the results found in \citet{Martin+17}}, but we also detect a non-negligible fraction QGs extending from low- to high-mass central systems.
	\change{As shown in \citetalias{Corcho-Caballero+22}, this population accounts for $\sim10$ per cent of the total number of nearby galaxies at $z\lesssim0.1$, and the characteristic timescale involved in the quenching of the SFR has a median value of $\sim 500$ Myr \citep[compatible with the estimates presented in][]{Weibel+22}.}
	In the \mass-\rfifth plane, they distribute between Ageing and Retired objects, relatively close to $\Sigma_{\rm Q}$.
	QGs feature a wide range of values of \vsigma, although we see that low-mass galaxies might be more rotationally-supported, while massive QGs are dominated by random motions.
	For any stellar mass, they systematically present higher metallicities than AGs.
	
	Using different estimates of the star formation activity is of paramount importance in order to identify Quenched galaxies.
	Any selection based on the \mass-\ssfr plane alone will always include a fraction of AGs and/or RGs, depending on the specific timescale of the adopted tracer.
	From to our AD classification scheme, we conclude that quenching in the local Universe is an ubiquitous processes, compatible with the idea of more than one channel acting at different regimes, but more frequent in dense environments.
	Given the physical trends found in this work, and the results from \citet{Corcho-Caballero+22}, we conclude that our satellite QG population is compatible with a post-starburst scenario driven by environmental processes such as ram pressure stripping, where a burst of star formation followed by a sudden drop of SFR led to a compact and metal-rich phase in most QGs.
	On the other hand, central QGs are compatible with the ``mass quenching'' scenario \citep{Peng+10}, although the mass-size relation of central QGs points toward ``morphological quenching'' \change{\citep{Martig+09, Gensior+20}}.
	In order to determine the rate of quenching along cosmic time and the build up of the Retired population, we suggest that a detailed analysis of the AD populations using larger, deep, and complete spectroscopic samples must be carried out.
	
	\section*{Acknowledgements}
	
	PC and YA thank financial support provided by grant PID2019-107408GB-C42 of the Spanish State Research Agency (AEI/10.13039/501100011033).
    A significant fraction of this work has been carried out during a long-term research visit of YA to ICRAR-UWA funded under the mobility scheme \textit{Ayudas para la Recualificaci\'on del Profesorado Universitario} of the UAM.
	SFS thanks the support by the PAPIIT-DGAPA IG100622 project, based on data obtained from the ESO Science Archive Facility.
	LC acknowledges support from the Australian Research Council Discovery Project and Future Fellowship funding schemes (DP210100337, FT180100066).
	This research made extensive use of the python packages NumPy,\footnote{https://numpy.org/} Matplotlib,\footnote{https://matplotlib.org/} \citep{Hunter:2007} and Astropy,\footnote{http://www.astropy.org} \citep{astropy:2013, astropy:2018}.
	
	Funding for the Sloan Digital Sky Survey IV has been provided by the Alfred P. Sloan Foundation, the U.S. Department of Energy Office of Science, and the Participating Institutions. 
	
	SDSS-IV acknowledges support and resources from the Center for High Performance Computing  at the University of Utah. The SDSS website is www.sdss.org.
	
	SDSS-IV is managed by the Astrophysical Research Consortium for the Participating Institutions of the SDSS Collaboration including the Brazilian Participation Group, the Carnegie Institution for Science, Carnegie Mellon University, Center for Astrophysics | Harvard \& Smithsonian, the Chilean Participation Group, the French Participation Group, Instituto de Astrof\'isica de Canarias, The Johns Hopkins University, Kavli Institute for the Physics and Mathematics of the Universe (IPMU) / University of Tokyo, the Korean Participation Group, Lawrence Berkeley National Laboratory, Leibniz Institut f\"ur Astrophysik Potsdam (AIP),  Max-Planck-Institut f\"ur Astronomie (MPIA Heidelberg), Max-Planck-Institut f\"ur Astrophysik (MPA Garching), Max-Planck-Institut f\"ur Extraterrestrische Physik (MPE), National Astronomical Observatories of China, New Mexico State University, New York University, University of Notre Dame, Observat\'ario Nacional / MCTI, The Ohio State University, Pennsylvania State University, Shanghai Astronomical Observatory, United Kingdom Participation Group, Universidad Nacional Aut\'onoma de M\'exico, University of Arizona, University of Colorado Boulder, University of Oxford, University of Portsmouth, University of Utah, University of Virginia, University of Washington, University of Wisconsin, Vanderbilt University, and Yale University.SDSS is managed by the Astrophysical Research Consortium for the Participating Institutions of the SDSS Collaboration including the Brazilian Participation Group, the Carnegie Institution for Science, Carnegie Mellon University, Center for Astrophysics | Harvard \& Smithsonian (CfA), the Chilean Participation Group, the French Participation Group, Instituto de Astrofísica de Canarias, The Johns Hopkins University, Kavli Institute for the Physics and Mathematics of the Universe (IPMU) / University of Tokyo, the Korean Participation Group, Lawrence Berkeley National Laboratory, Leibniz Institut für Astrophysik Potsdam (AIP), Max-Planck-Institut für Astronomie (MPIA Heidelberg), Max-Planck-Institut für Astrophysik (MPA Garching), Max-Planck-Institut für Extraterrestrische Physik (MPE), National Astronomical Observatories of China, New Mexico State University, New York University, University of Notre Dame, Observatório Nacional / MCTI, The Ohio State University, Pennsylvania State University, Shanghai Astronomical Observatory, United Kingdom Participation Group, Universidad Nacional Autónoma de México, University of Arizona, University of Colorado Boulder, University of Oxford, University of Portsmouth, University of Utah, University of Virginia, University of Washington, University of Wisconsin, Vanderbilt University, and Yale University.
	
	The SAMI Galaxy Survey is based on observations made at the Anglo-Australian Telescope. The Sydney-AAO Multi-object Integral field spectrograph (SAMI) was developed jointly by the University of Sydney and the Australian Astronomical Observatory. The SAMI input catalogue is based on data taken from the Sloan Digital Sky Survey, the GAMA Survey and the VST ATLAS Survey. The SAMI Galaxy Survey website is http://sami-survey.org/. The SAMI Galaxy Survey is supported by the Australian Research Council Centre of Excellence for All Sky Astrophysics in 3 Dimensions (ASTRO 3D), through project number CE170100013, the Australian Research Council Centre of Excellence for All-sky Astrophysics (CAASTRO), through project number CE110001020, and other participating institutions.
	
	\section*{Data Availability}
	
	The data underlying this article are publicly available at https://www.sdss.org/dr17/manga/ for MaNGA and http://sami-survey.org for the
	SAMI surveys, respectively.
	Additional data generated by the analyses in this
	work are available upon request to the corresponding author.

	
	
	\bibliographystyle{mnras}
	\bibliography{bibliography} 

	
	
	\appendix

\section{MaNGA and SAMI statistics}
Table~\ref{tab:percentiles} includes the median and dispersion values (estimated with the 15 and 84 percentiles) derived from each property for \man and \sam samples.
\begin{table*}
	\centering
	\begin{tabular}{c|c|c|c|c}
		\hline 
		& Ageing (MaNGA|SAMI) & Undet. (MaNGA|SAMI) & Quenched (MaNGA|SAMI) & Retired (MaNGA|SAMI)\\ 
		\hline \vspace{5pt} 
		$D_n(4000)$ & $1.27^{+0.26}_{-0.15}$ | $1.20^{+0.27}_{-0.13}$ &$1.26^{+0.14}_{-0.18}$ | $1.17^{+0.16}_{-0.16}$ &$1.57^{+0.07}_{-0.10}$ | $1.54^{+0.08}_{-0.14}$ &$1.80^{+0.06}_{-0.08}$ | $1.79^{+0.07}_{-0.08}$\\ \vspace{5pt} 
		$\rm EW(H\alpha)~(\angstrom)$ & $13.46^{+16.54}_{-10.53}$ | $16.50^{+20.01}_{-12.21}$ &$0.17^{+2.29}_{-0.53}$ | $1.78^{+5.92}_{-2.18}$ &$-1.84^{+0.42}_{-0.30}$ | $-1.73^{+0.30}_{-0.29}$ &$-1.05^{+0.60}_{-0.39}$ | $-1.33^{+0.73}_{-0.32}$\\ \vspace{5pt} 
		$\log_{10}(M_\star/\rm M_\odot)$ & $10.25^{+0.73}_{-0.68}$ | $9.77^{+0.73}_{-1.06}$ &$9.95^{+1.16}_{-0.70}$ | $8.97^{+0.93}_{-0.67}$ &$9.75^{+0.62}_{-0.37}$ | $9.78^{+0.51}_{-0.47}$ &$11.01^{+0.45}_{-0.68}$ | $10.54^{+0.45}_{-0.42}$\\ \vspace{5pt} 
		$\log_{10}(R_{50}/\rm kpc)$ & $0.55^{+0.25}_{-0.26}$ | $0.51^{+0.25}_{-0.31}$ &$0.51^{+0.30}_{-0.33}$ | $0.25^{+0.28}_{-0.27}$ &$0.24^{+0.25}_{-0.15}$ | $0.37^{+0.28}_{-0.23}$ &$0.60^{+0.38}_{-0.37}$ | $0.52^{+0.25}_{-0.23}$\\ \vspace{5pt} 
		$\frac{R_{90}}{R_{50}}$ & $2.55^{+0.44}_{-0.35}$ | $2.24^{+0.48}_{-0.27}$ &$2.66^{+0.54}_{-0.42}$ | $2.26^{+0.45}_{-0.36}$ &$2.87^{+0.33}_{-0.28}$ | $2.64^{+0.79}_{-0.36}$ &$3.20^{+0.26}_{-0.37}$ | $3.18^{+0.44}_{-0.43}$\\ \vspace{5pt} 
		$\frac{v}{\sigma}$ & $0.45^{+0.16}_{-0.15}$ | $0.60^{+0.27}_{-0.29}$ &$0.41^{+0.24}_{-0.16}$ | $0.38^{+0.47}_{-0.16}$ &$0.43^{+0.19}_{-0.21}$ | $0.33^{+0.22}_{-0.14}$ &$0.21^{+0.14}_{-0.11}$ | $0.29^{+0.21}_{-0.16}$\\ \vspace{5pt} 
		$\log_{10}(Z_{\rm ssp}/Z_\odot)$ & $-0.27^{+0.23}_{-0.44}$ | $-0.82^{+0.70}_{-0.70}$ &$-0.28^{+0.17}_{-0.54}$ | $-1.14^{+0.58}_{-0.54}$ &$-0.19^{+0.12}_{-0.12}$ | $-0.26^{+0.18}_{-0.36}$ &$0.00^{+0.08}_{-0.10}$ | $-0.01^{+0.14}_{-0.12}$\\ \vspace{5pt} 
		$\log_{10}\left(\frac{\Sigma_5}{\rm Mpc^{-2}}\right)$ & $-0.83^{+0.66}_{-0.61}$ | $0.01^{+0.86}_{-0.65}$ &$-0.86^{+0.94}_{-1.14}$ | $0.34^{+0.96}_{-0.76}$ &$-0.18^{+0.83}_{-0.94}$ | $1.26^{+0.47}_{-0.68}$ &$-0.60^{+0.96}_{-1.40}$ | $1.00^{+0.66}_{-1.23}$
		\\ \hline
	\end{tabular}
	\caption{Median values of each property ($\pm$ 1 sigma estimated with the 16 and 84 percentiles) for Ageing, Undetermined, Quenched and Retired populations, corresponding to the values shown in Fig.~\ref{fig:manga_cornerplot}, and Fig.~\ref{fig:sami_cornerplot} for \man and \sam samples, respectively. }
	\label{tab:percentiles}
\end{table*}

\bsp	
\label{lastpage}

\end{document}